\documentclass[journal]{IEEEtran}
\pdfoutput=1
\usepackage{times}
\usepackage{graphicx}
\usepackage{amsmath}
\usepackage{lettrine}
\usepackage{setspace}
\usepackage{epstopdf}
\usepackage{subfigure}
\usepackage[usenames, dvipsnames]{color}

\begin{document}

\title{Exploiting OxRAM Resistive Switching for Dynamic Range Improvement of CMOS Image Sensors}

\author{Ashwani Kumar,~\IEEEmembership{Student Member,~IEEE,} Mukul Sarkar,~\IEEEmembership{Member,~IEEE,}
        and Manan Suri,~\IEEEmembership{Member,~IEEE,}
  
\thanks{Authors are with the Department of Electrical Engineering, Indian Institute of Technology, Delhi, New Delhi 110016, India (e-mail: manansuri@ee.iitd.ac.in) }
\vspace{-1em}
}

\maketitle

\begin{abstract}
We present a unique application of OxRAM devices in CMOS Image Sensors (CIS) for dynamic range (DR) improvement. We propose a modified 3T-APS (Active Pixel Sensor) circuit that incorporates OxRAM in 1T-1R configuration. DR improvement is achieved by resistive compression of the pixel output signal through autonomous programming of OxRAM device resistance during exposure. We show that by carefully preconditioning the OxRAM resistance, pixel DR can be enhanced. Detailed impact of OxRAM SET-to-RESET and RESET-to-SET transitions on pixel DR is discussed. For experimental validation with specific OxRAM preprogrammed states, a 4 Kb 10 nm thick HfOx (1T-1R) matrix was fabricated and characterized. Best case, relative pixel DR improvement of $\sim$ 50 dB was obtained for our design. 
\end{abstract}


\IEEEpeerreviewmaketitle

\section{Introduction}
\begin{figure}
\center
\includegraphics[width=70mm,height=60mm]{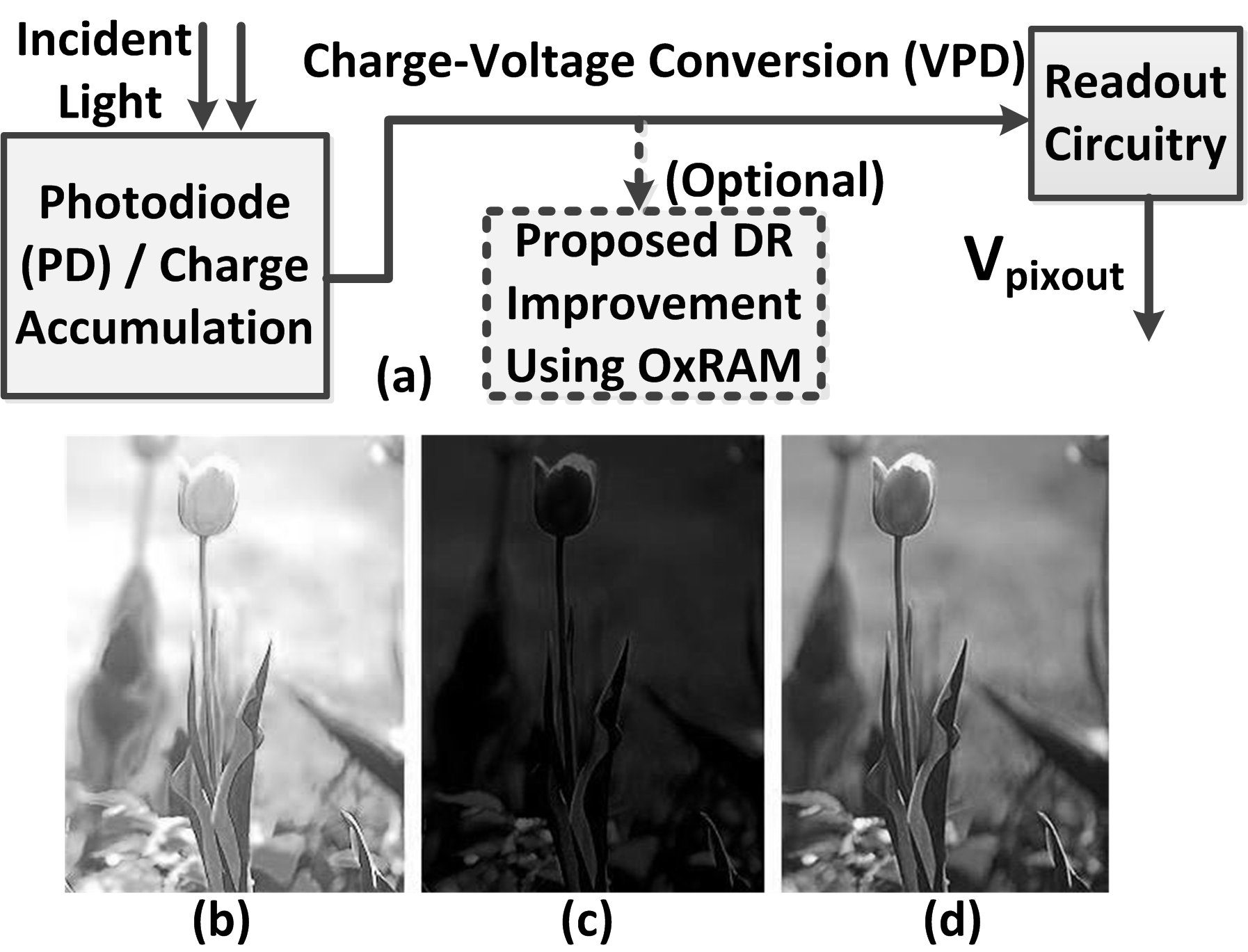}
\caption{(a) Schematic of CIS pixel with proposed additional OxRAM block for dynamic range improvement. Representative grayscale illustrations showing impact of (b) bright, (c) dim, and (d) normal, -light exposure.} 
\vspace{-1em}
\label{bd1}
\end{figure}

\IEEEPARstart{C}{MOS} Image Sensors (CIS) are the heart of modern day imaging systems. Block diagram of a typical CIS pixel, is shown in Fig.~\ref{bd1}(a) along with the proposed optional block for dynamic range (DR) enhancement. Most CIS operate photodiode (PD) in charge accumulation mode~\cite{coc} and PD is initially reset to a high potential~\cite{coc}. Incident light on PD leads to generation of charges (electrons), that are collected at the photodiode's capacitor (C$_{PD}$). When incident light is low, few electrons are generated and collected at C$_{PD}$, leading to a small voltage drop at PD. On the other hand, bright light generates large number of charges and thus a larger voltage drop is realized at PD. 

Pixel DR is directly proportional to the ratio of maximum to minimum detectable exposure current~\cite{ael}. DR is also defined as ratio of maximum output voltage swing to minimum valid detectable output (above noise in dark).
Maximum output voltage swing of the pixel can either be limited by the saturation of the readout circuitry or by the PD Full Well Capacity (FWC)~\cite{Voswing,fwc}.
Whereas, minimum detectable exposure is limited by the sensitivity and noise floor at (PD) or limitation of readout circuitry ~\cite{jn,ael}.
High dynamic range (HDR) is one of the most desirable characteristic of any imaging system~\cite{jn}, which enables the detection of a broad range of illumination.
However, most CIS pixels suffer from low dynamic range (DR). 
Fig.~\ref{bd1} (b)-(d) shows sample images (in grayscale for illustrative purpose) in bright-, dim-, and normal- lighting conditions, when a pixel DR is not optimized. Several solutions exist for DR improvement, such as - noise reduction ~\cite{negfeed}, managing the sensitivity of PD ~\cite{hyne}, use of logarithmic pixel response ~\cite{coc},~\cite{lin_log}, two-stage integration under high illumination ~\cite{XDR}, stepped reset-gate voltage technique ~\cite{stp_rst}, reconfigurable modes of operation ~\cite{reconf_modes} and others ~\cite{quant_ana}. These solutions come either at the cost of increased silicon/pixel area, or added circuit complexity, or a combination of both. 

In recent work ~\cite{ash}, we showed that ‘capacitive’ properties of OxRAM devices (Metal oxide based resistive nonvolatile memory) ~\cite{cv1,cv2} can be exploited for pixel DR improvement in bright-light conditions. In particular, we show in ~\cite{ash} how non-linear CV of HfOx based OxRAM, can be used to enhance the DR of a 4T CMOS pixel by a factor of ~ 2.45 X. The OxRAM in our previous approach ~\cite{ash} is essentially used as a high density non-linear MIMCAP, always programmed in a strong HRS state with no resistive switching event taking place during exposure or pixel operation.

In this paper, we throughly extend our previous capactive-based study, by showing a unique and first of its kind solution that exploits resistive-switching property of OxRAM devices for enhancing the operating DR in a CMOS 3T pixel circuit. The proposed effect, mainly depends on the modulation of OxRAM Conductive Filament (CF). This work opens the door for a new exciting non-storage (compression/imaging) application of emerging OxRAM technology. High integration density, low fabrication cost, CMOS compatibility ~\cite{wong}, full back-end integration, make OxRAM a suitable candidate for use in imaging/pixel applications.

\section{OxRAM Basics and Proposed Concept}
\subsection{OxRAM Basics}
OxRAM (or RRAM) is a two-terminal MIM-type (metal-insulator-metal) structure sandwiching a thin active insulating metal oxide layer between top electrode (TE) and bottom electrode (BE) as shown in Fig.~\ref{concept1} . 
 
The active oxide layer exhibits reversible non-volatile resistive switching behavior on application of appropriate voltage/current across the device terminals. Formation of a conductive filament (CF), comprising of oxygen vacancies and defects in the active layer, switches the device to a low resistance state (LRS / $R_{SET}$), while rupturing of the CF switches the device to a high-resistance state (HRS / $R_{RESET}$ ) ~\cite{wong}. Resistive behavior of OxRAM devices has been well characterized~\cite{wong}, and modeled ~\cite{wong2D,dam1,dam2} for HfOx and several other metal oxide stacks. 
 
 OxRAM resistance switching and programmability has been widely proposed for use in conventional non-volatile storage applications (replacement of Flash), and advanced applications such as- neuromorphic computing~\cite{garb}, machine-learning~\cite{man} and embedded design~\cite{vian}. In this work we extend the application scope to the domain of hybrid CMOS-OxRAM imaging pixels. In particular, OxRAM devices with HRS ($\sim$ G$\Omega$), and fast switching (sub-microsecond) are ideal for the proposed pixel application. As explained in detail in Section III, and Section IV, OxRAM's high $R_{RESET}$ and fast (sub-microsecond) switching are desirable as these help to restrict the current flow from pixel to ground during exposure, thus enabling the manipulation of pixel DR during operation. In literature, several demonstrations of state-of-the-art HfOx devices with such desired attributes like high $R_{RESET}$ and fast switching, exist ~\cite{si_hf1.25n,gaa_hf200p,nisili_hf1n,tao_hf1n}. 
 
\subsection{Proposed Concept for DR Improvement}
 
In our proposed solution, an additional block, consisting of an emerging non-volatile resistive memory device i.e. OxRAM is added for DR enhancement as shown in Fig.~\ref{bd1}(a). The OxRAM (in 1T-1R configuration) is connected to the pixel, such that during exposure, pixel signal (VPD) gets applied at one of the OxRAM electrode. Prior to exposure, the OxRAM device is pre-programmed to a specific initial resistance level (either in LRS and HRS). At the start of exposure, application of VPD signal may modulate the conductive filament (CF), and thus the resistance of the OxRAM device. Modulation of the CF during the exposure, and OxRAM final resistance at the end of the exposure, will be a function of (i) OxRAM initial resistance state, (ii) VPD signal (i.e. the effective voltage drop across the OxRAM terminals arising due to VPD), and the (iii) I-V switching characteristics of the OxRAM device (the currents/voltages needed to switch the device to a particular state). 

\begin{figure}
\center
\includegraphics[width=60mm,height=55mm]{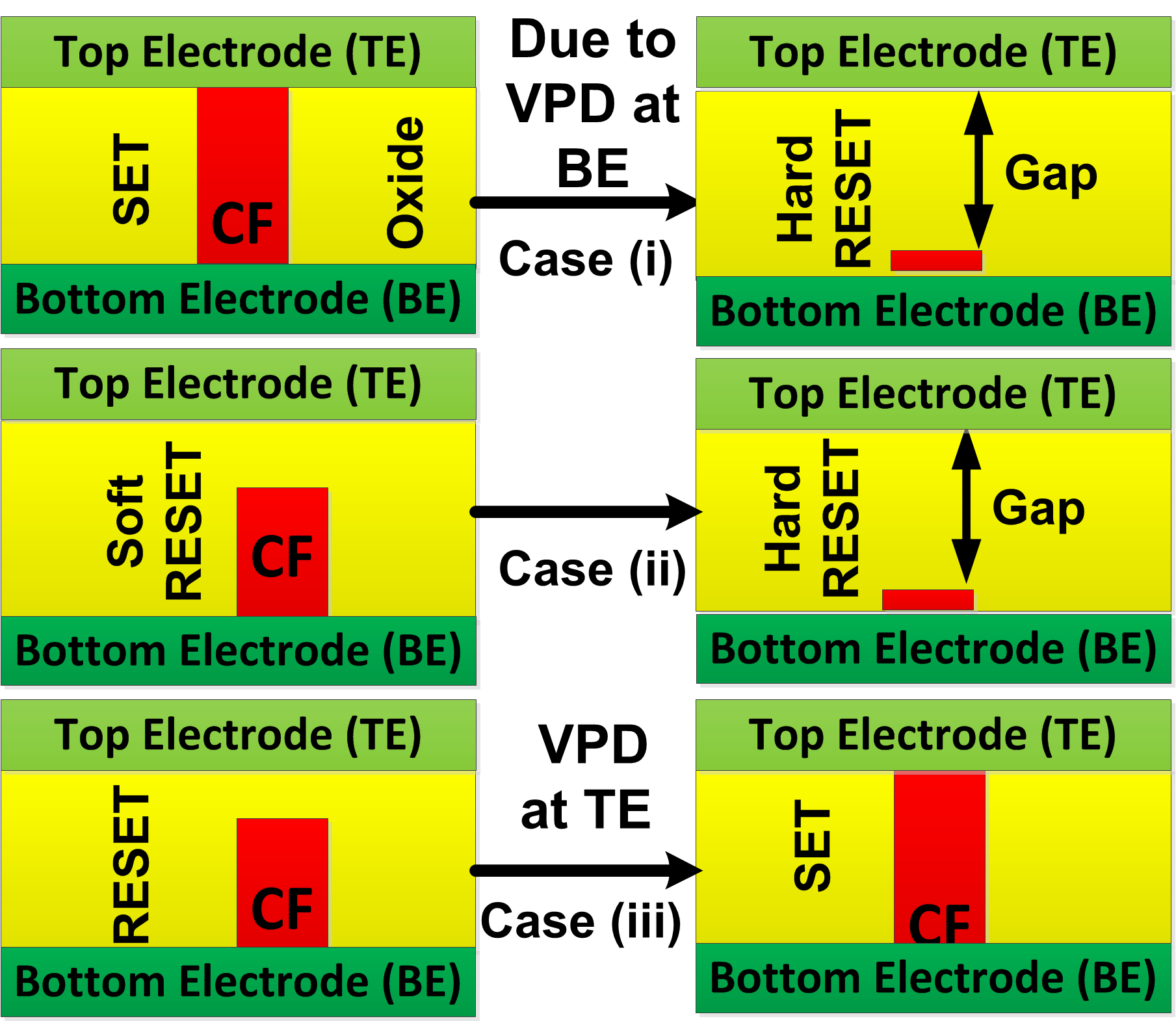}
\caption{Conceptual diagram showing different cases of OxRAM CF modulation exploited in our proposed methodology. VPD discharge/evolution through the OxRAM is manipulated by the OxRAM CF modulation and the current flowing through it.} 
\label{concept1}
\end{figure}

Depending on specific I-V characteristic, OxRAM initial resistance can be programmed such that it switches to a particular resistance state when a certain VPD drop occurs across the device terminals. Further, different type of switching events (LRS/HRS) and their corresponding CF modulations, will determine the nature of VPD signal evolution (or VPD fall) during the exposure duration. For instance, formation of a stronger filament at exposure (i.e. switching to strong SET state), would lead to a faster drain of VPD signal. Thus the evolution of VPD can be changed to some extent by indirectly controlling the nature of switching event that the OxRAM device undergoes during the exposure. VPD can be manipulated, such that it is  compressed with different gain factors (GF) during the exposure. This helps to control/tune the overall DR of the pixel, as shown in Section III and IV.
Fig.~\ref{concept1} illustrates three different switching cases or events of OxRAM resistance (or CF) modulation that we studied for pixel DR enhancement. In case(i), the OxRAM undergoes SET to RESET transition. In case (ii), the OxRAM undergoes a soft-RESET to hard-RESET transistion. Whereas in case (iii), the OxRAM undergoes a RESET to SET transition. Detailed impact of each of these transitions on the pixel output is explained in Section III and IV.   

\vspace{-1em}

\section{Proposed Hybrid CMOS-OxRAM Pixel Circuit}
All circuit simulations presented in this paper, were performed in Cadence Virtuoso environment using OxRAM compact model presented in ~\cite{wong2D} (10 nm HfOx device), and UMC 180 nm CMOS technology-node. We choose 180 nm node for CMOS, as pixel performance begins to degrade with aggressive scaling due to effects such as increased gate leakage, sub-vt leakage, dark current, and shot noise ~\cite{pixel_scaling}. OxRAM technology may have fewer constraints if it’s fully integrated in the pixel backend, thus we choose an OxRAM device for our study with a smaller feature size compared to CMOS. As discussed in Section IV, large dimensions of OxRAM electrodes (contact area) will give rise to a higher value of C$_{POX}$ (OxRAM intrinsic capacitance), which is detrimental for the proposed application specifically for dim-light correction. In all simulations, a  time duration of 0.5 $\mu$s is assigned for PD reset operation before the start of exposure. Light exposure is modeled with current source (I$_{exp}$), using a fixed exposure period of 9.5 $\mu$s. Lower values of I$_{exp}$ correspond to dim-light, while higher values correspond to bright-light stimuli.

\subsection{Working of 3 Transistor-Active Pixel Sensor (APS) with and without OxRAM}

\begin{figure}
\center
\includegraphics[width=90mm,height=90mm]{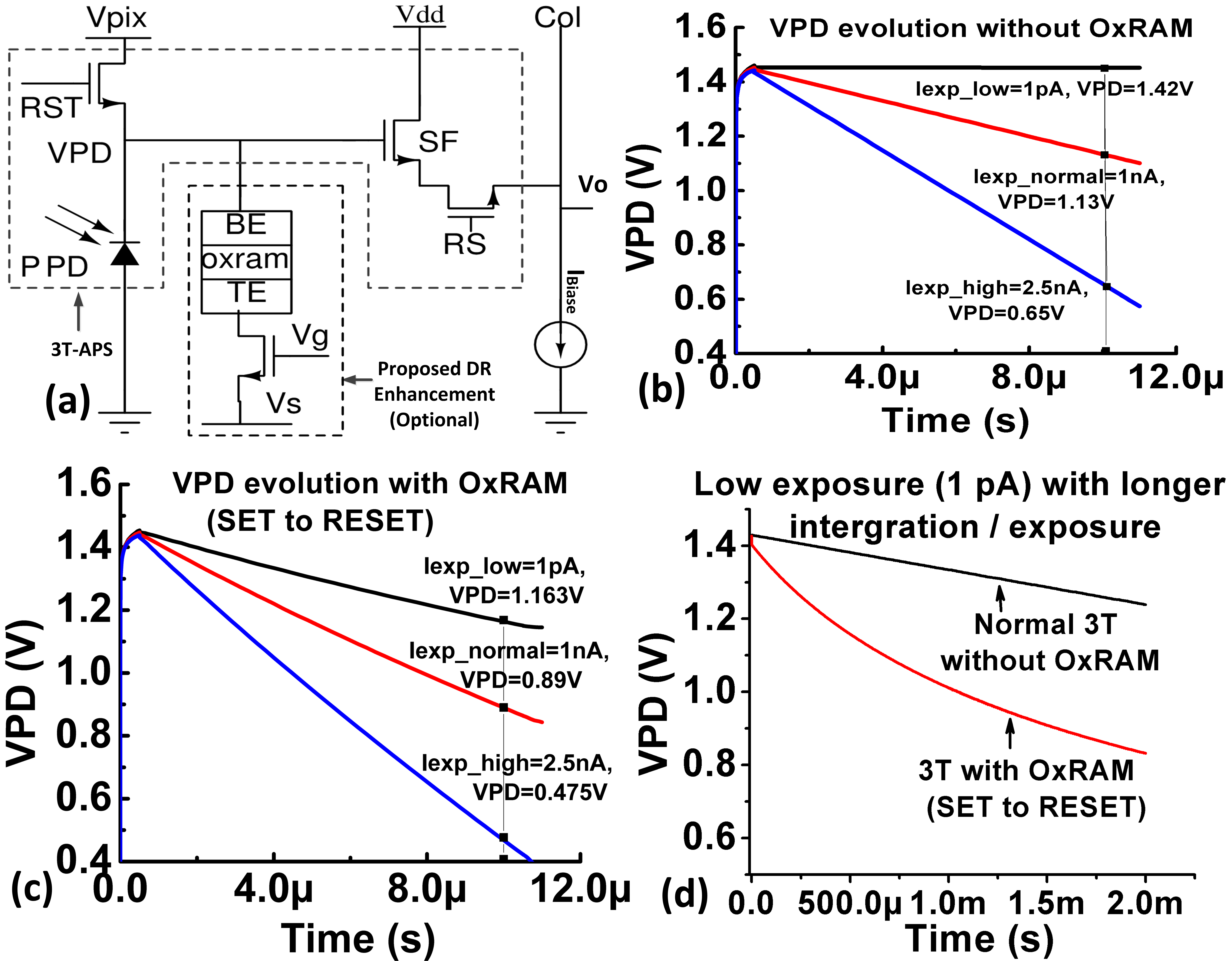}
\caption{(a) 3T-APS with proposed OxRAM based DR enhancement module. Simulated VPD evolution for high-, normal- and low- exposure cases in- (b) 3T-APS, (c) 3T-APS with OxRAM (hybrid pixel). R$_{SET}$ = 1.25 M$\Omega$; Vreset = 1.42 V; Ireset = $\sim$ 11 $\mu$A; Treset = $\sim$ 510 ns; R$_{RESET}$ = $\sim$ 60 G$\Omega$ (Vread = 0.1 V). (d) Case with low exposure current and longer exposure time.} 
\vspace{-1em}
\label{pix1}
\end{figure}

Fig.~\ref{pix1}(a) shows 3T-APS with the optional OxRAM block for `Operating-DR' improvement. In our case, Operating-DR corresponds to the estimated DR of the 3T APS, considering the limitations of the readout circuitry, assuming a maximum readable output voltage swing of 0.85 V (operating range).  

In 3T-APS (without OxRAM), initially VPD is at the Vrst level (1.42 V). VPD evolution is a direct consequence of the light exposure on PD. Fig.~\ref{pix1}(b) shows the simulated evolution of VPD signal under three different exposure conditions (high, normal and low) for the pixel without the OxRAM. For the low exposure  case (I$_{exp}$ = 1 pA), VPD final value (at the end of exposure) does not remain in the operating range and stays close to initial fixed reset potential (Vrst). For the high exposure case (I$_{exp}$ = 2.5 nA), VPD remains in the operating range. 

We simulated the same exposure cases using the proposed hybrid 3T-pixel with OxRAM as shown in Fig.~\ref{pix1}(c). Controlling and addressing a specific pixel in the entire array would also require use of peripheral circuitry similar to the one used for CIS or RRAM memory arrays~\cite{ox_prog}. For pre-programming OxRAM device in an individual pixel, first transistor RS should be turned off, disconnecting the read circuitry. To Set the OxRAM- Vg, RST, and Vs are raised to a high potential and Vpix is kept at low potential. To Reset the OxRAM- Vg, RST, and Vpix are raised to a high potential and Vs is grounded.
OxRAM device in case of Fig.~\ref{pix1}(c), was pre-programmed to SET state and then final VPD value for low exposure, remains in the operating range unlike the case in Fig.~\ref{pix1}(b).
%\textcolor{blue}{Here, OxRAM switches from pre-programming set state resistance (R$_{SET}$ = 1.25 M$\Omega$) to the reset state (R$_{RESET}$ = $\sim$ 60 G$\omega$). During the simulation, initial VPD value (1.42 V) at the BE of OxRAM   }
For high exposure, VPD remains near the lower verge of the operating range. Fig.~\ref{pix1}(d), shows the impact of longer exposure duration, ($\sim$ 2 ms) with low exposure current (1 pA).

\begin{figure}
\center
\includegraphics[width=70mm,height=50mm]{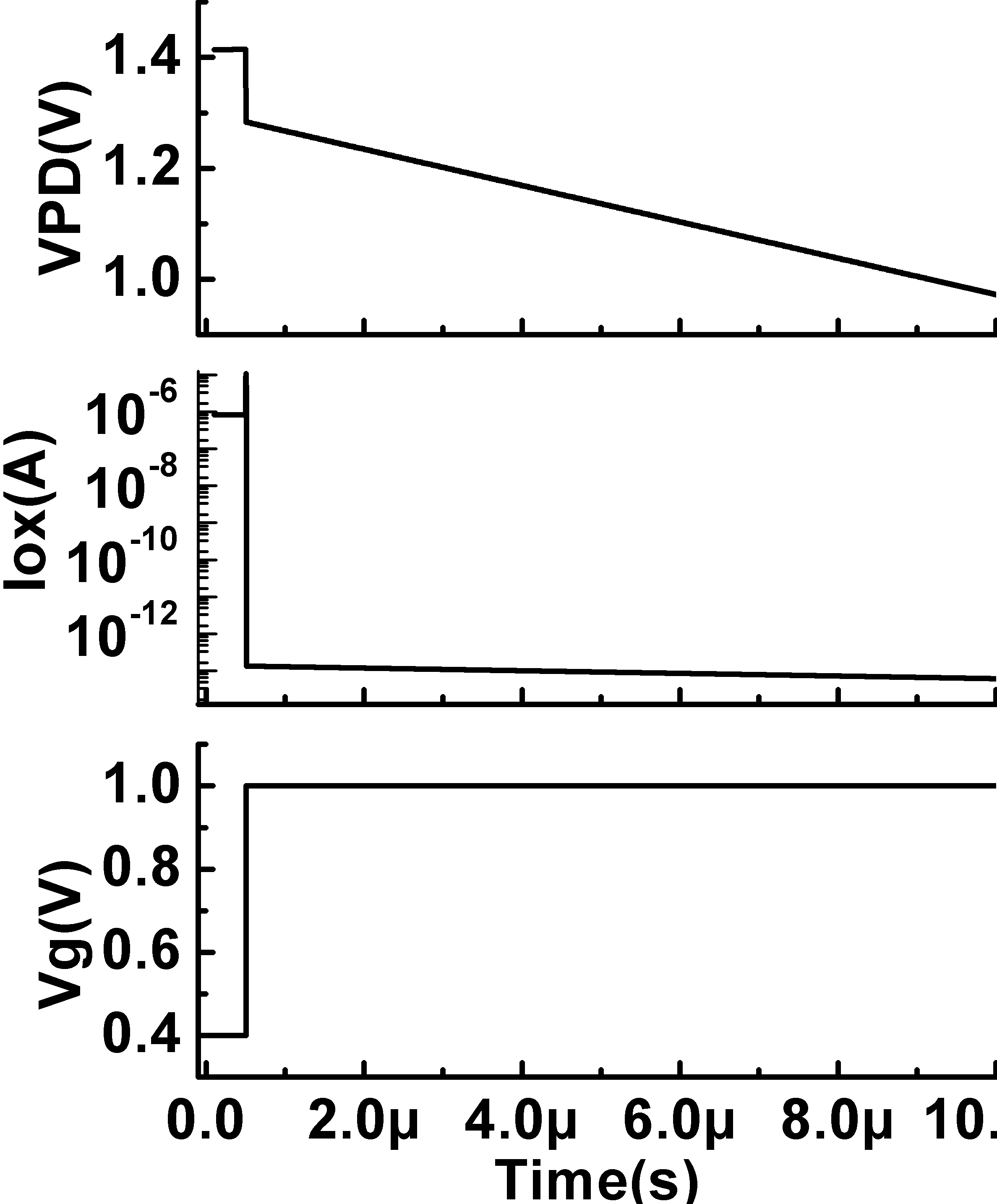}
\caption{Simulated transient waveforms of Vg, VPD, and Iox during one complete exposure operation. Peak in the current waveform illustrates OxRAM Set to Reset switching event.} 
\vspace{-1em}
\label{Vg_tran}
\end{figure}

Thus, the use of OxRAM modulates the fall/evolution of the VPD signal during the exposure. In particular, for the low-exposure cases, the OxRAM helps in retaining the final VPD level in the operating range, virtually extending the operating-DR of the system. When initial state of OxRAM is SET, the strong CF may act as a path and lead to a sudden fall in VPD during exposure (i.e. before the OxRAM actually switches to Reset state, or the CF is not yet erased). Such abrupt VPD fall can be controlled by application of a suitable gate voltage signal (Vg) at the OxRAM selector device. Transient waveforms (Vg, VPD, Iox), illustrating such suitable gate control to prevent sudden VPD fall are shown in Fig. ~\ref{Vg_tran}.
Section IV discusses in detail the current based mechanism (i.e. current flowing through the evolving CF of the OxRAM) that enables this virtual DR enhancement.  

\begin{figure}
\center
\includegraphics[width=90mm,height=90mm]{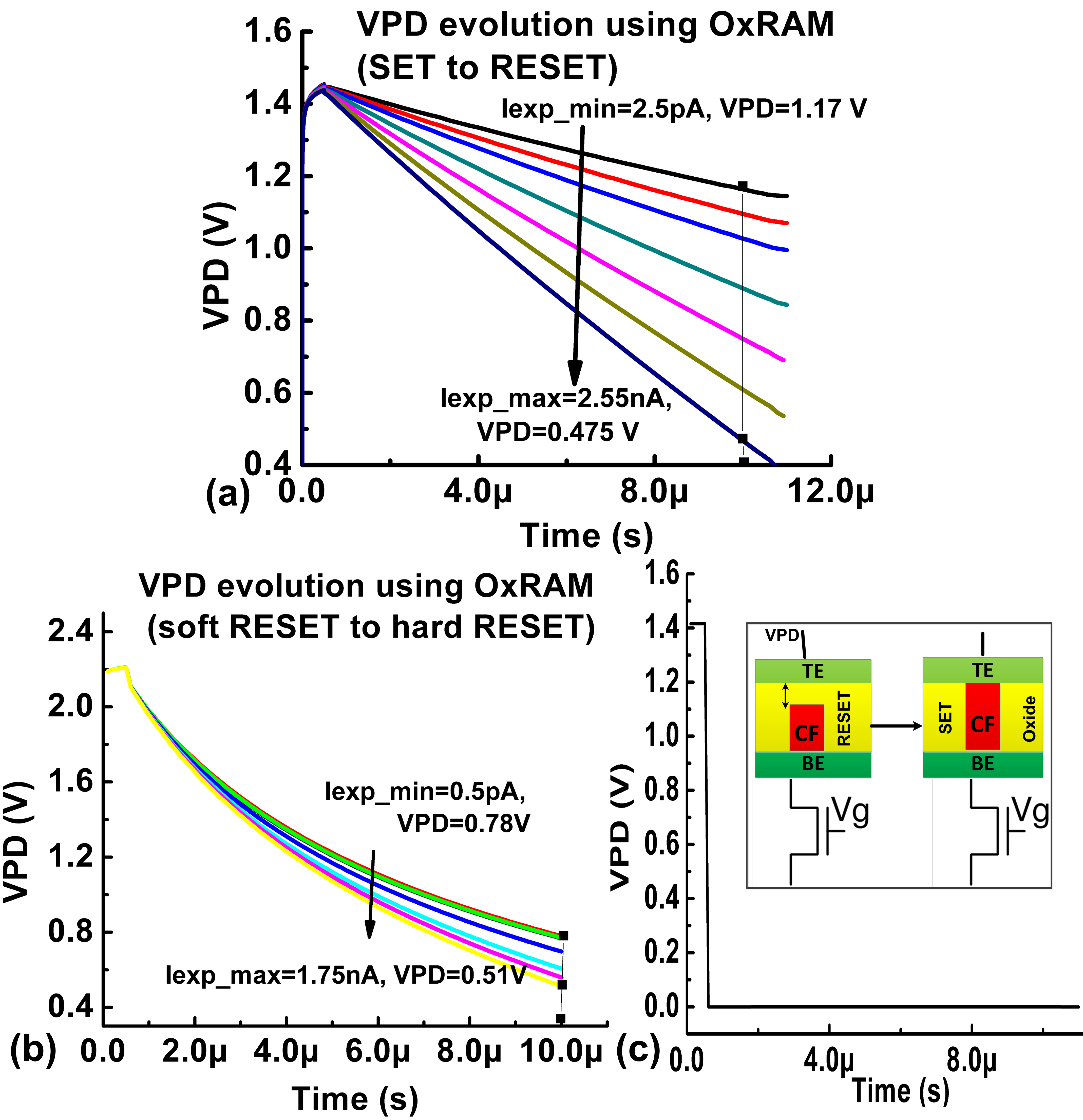}
\caption{VPD evolution of hybrid pixel for different exposure conditions, corresponding to different OxRAM switching events: (a) SET to RESET corresponding to case(i), (b) soft-RESET to hard-RESET, corresponding to case (ii), and (c) RESET to SET, corresponding to case (iii). Inset (c) VPD falls abruptly due to formation of a strong CF.} 
\vspace{-1em}
\label{vpds1}
\end{figure}

\begin{figure}
\center
\includegraphics[width=70mm,height=60mm]{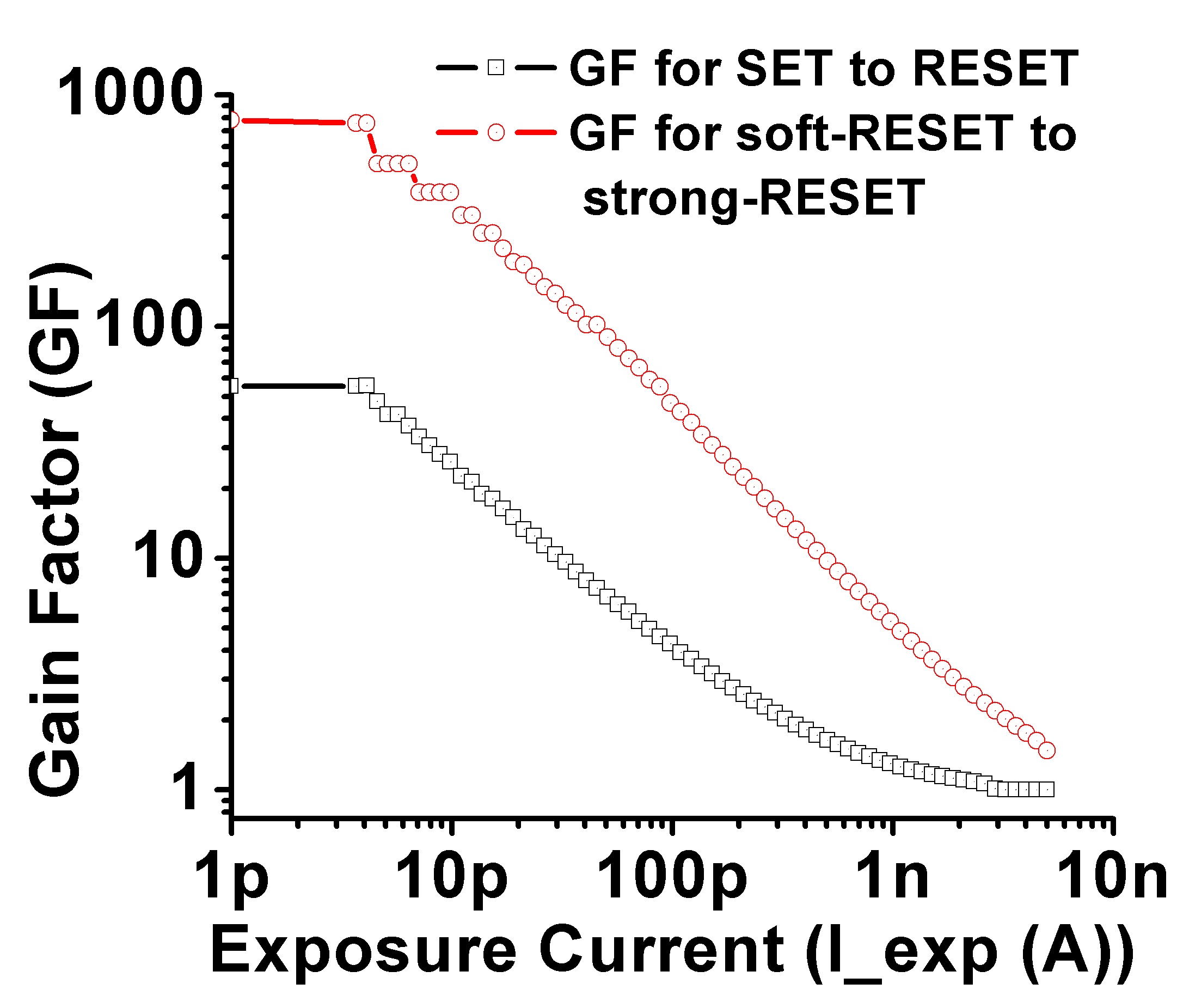}
\caption{Gain Factor (GF) curves of hybrid pixel for two different switching cases of OxRAM.} 
\vspace{-1em}
\label{GF1}
\end{figure}

\subsection{Physical Impact of OxRAM Switching Event Type}
We simulated the hybrid pixel for three different OxRAM switching cases, shown in Fig.~\ref{concept1}. The exposure current was varied over a wide range (100 fA to 10 nA) to emulate a large number of stimuli conditions.

Fig.~\ref{vpds1}(a), shows the VPD evolution for full exposure duration for case (i), i.e. OxRAM transition from SET to RESET state. In this case, OxRAM BE was connected to VPD node and TE to its selector/control device (transistor) as shown in Fig.~\ref{pix1}(a). The OxRAM was first initialized to a SET state (pre-exposure). At the start of the exposure, a high positive potential exists at the OxRAM BE thus switching it from SET to RESET state. Thus, VPD evolution becomes a consequence of both exposure as well as OxRAM resistance states before and after the switching.  Post switching to RESET state the dependence of VPD evolution on conduction through the OxRAM diminishes as the CF is also disrupted. Final VPD values were observed to saturate at 1.17 V corresponding to any exposure below 2.5 pA. However, the use of OxRAM, as shown in Fig.~\ref{vpds1}(a), extends the range of exposure currents for which the final VPD voltage value remains in the detectable range.     
Case (ii) involves the transition of OxRAM from soft-RESET to hard-RESET state. Consequent VPD evolution is shown in Fig.~\ref{vpds1}(b).The OxRAM connections are identical as for case (i). OxRAM was initialized to a soft/partial-RESET state (pre-exposure). Due to initial ruptured CF of OxRAM, significant fraction of VPD drops across the ruptured region itself. This limits the amount of current that can flow during the exposure region through the device, thus inhibiting the device from going further in to a hard-RESET state. Thus, in order to achieve hard-RESET, we used higher VPD reset level (i.e. 2.2 V). With a higher reset level, the VPD evolution again became a consequence of both- exposure, and OxRAM resistance states pre- and post- switching. The detectable exposures are  illustrated in Fig.~\ref{vpds1}(b) with their corresponding VPD evolutions. In case (ii), VPD evolution follows a stronger exponential trend, as the OxRAM remains in the RESET state both pre- and post-exposure. The range of exposure current (0.5 pA to 1.75 nA), that gives final VPD values in the operating/readable range, also increases for case (ii) thus improving the operating-DR. 
Case (iii) corresponds to the OxRAM switching from RESET to SET state (Fig.~\ref{vpds1}(c)). VPD falls abruptly due to formation of a strong CF. Thus, no detectable VPD values can be obtained in this case. For RESET to SET switching (i.e. case iii), OxRAM TE is connected to PD node as shown in the inset of Fig.~\ref{vpds1}(c). Thus, DR improvement and DR tuneability was achievable for cases where the OxRAM switches to RESET state, or the CF is further erased compared to the OxRAM initial state.

Fig.~\ref{GF1} shows the gain factor (GF) of hybrid pixel for two different OxRAM switching cases. GF is calculated using (1) 
\begin{equation}
GF = \frac{Vrst - VPD2}{Vrst - VPD1}
\end{equation}
Where, Vrst is the fixed reset voltage level of PD. VPD1 corresponds to- poxt-exposure values for pixel without OxRAM, while VPD2 corresponds to post-exposure values for pixel with OxRAM.
GF remains fixed at a high value for low exposures. At low exposures, VPD1 remains close to Vrst while VPD2 is pulled to a lower value due to OxRAM switching.

\subsection{Impact of OxRAM Pre-programming}

\begin{figure}
\center
\includegraphics[width=70mm,height=90mm]{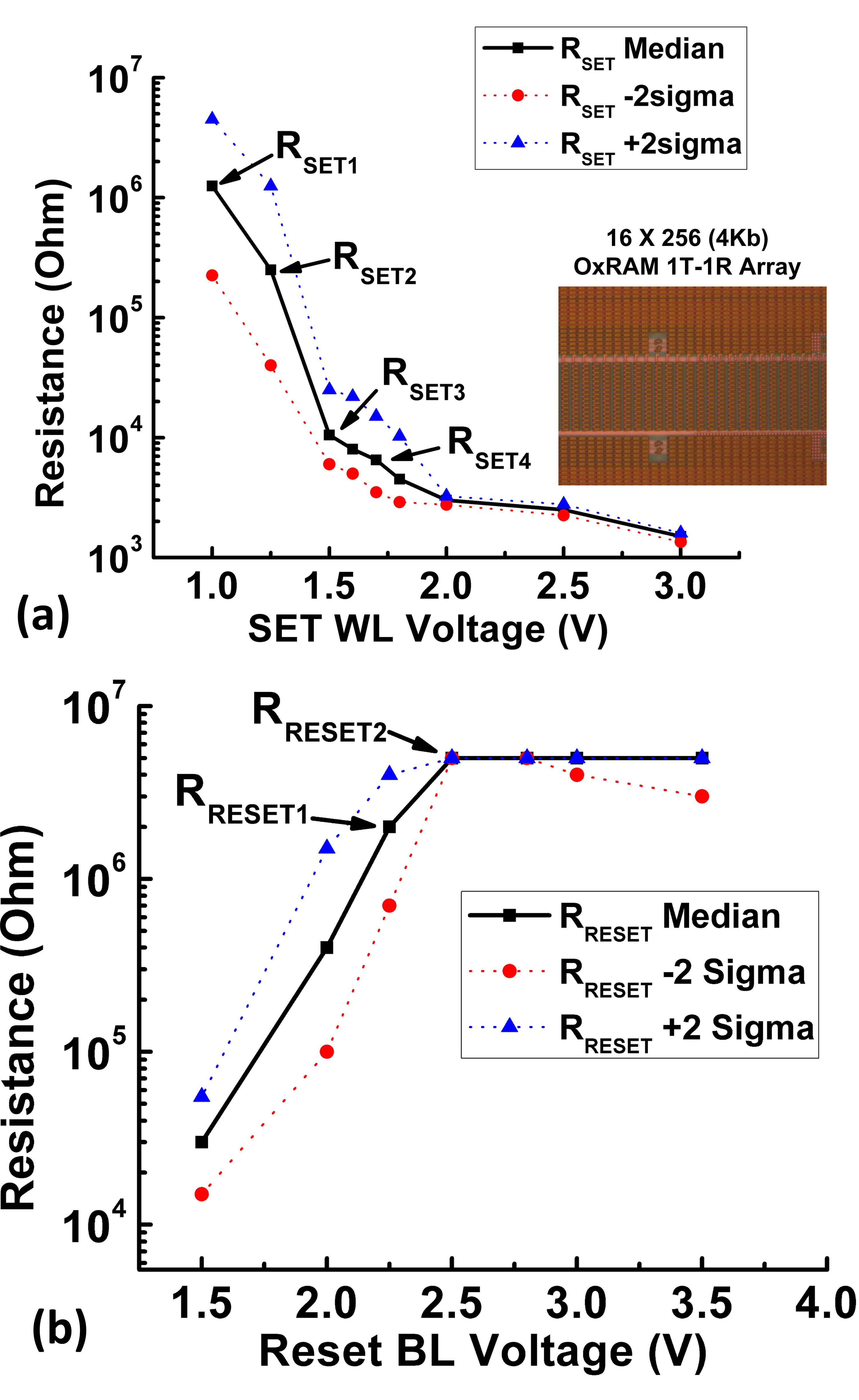}
\caption{Experimentally obtained distributions of resistance states on the 4 Kb OxRAM matrix: (a) SET state resistance levels (R$_{SET1- 4}$) after the RESET state (achieved using following conditions -  VWL = 3.2 V, VBL = 2.5 V, Tpulse = 100 ns). Inset shows optical micrograph of the fabricated matrix. (b)RESET state resistance levels R$_{RESET}$ after the SET state (achieved using following conditions -  VWL = 1.6 V, VBL = 2.5 V, Tpulse = 100 ns).  (Levels \textgreater 5 M$\Omega$ not visible due to read test bench limitation.)}
\vspace{-1em}
\label{rstates}
\end{figure}

\begin{figure}
\center
\includegraphics[width=70mm,height=90mm]{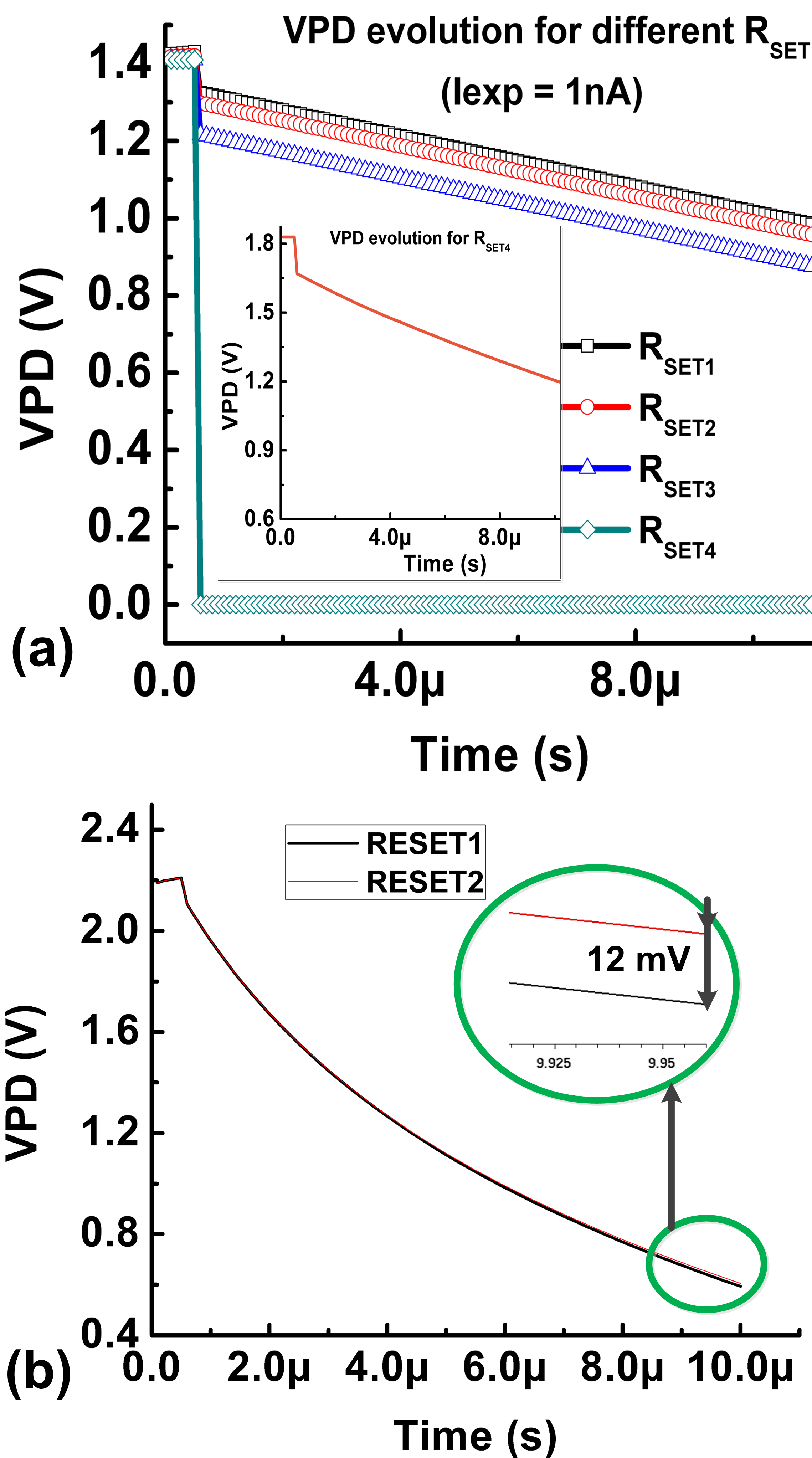}
\caption{Simulated VPD evolution of hybrid pixel for (a) 4 different pre-programmed $R_{SET}$ (R$_{SET1- 4}$). For high R$_{SET}$ (\textless 10 K$\Omega$) VPD falls
abruptly. Inset also shows simulated VPD for R$_{SET4}$ where VPD fall is not-abrupt. Here, VPD starting level is changed from 1.42 V to 1.8 V. (b) VPD evolution for R$_{RESET1}$ and R$_{RESET2}$. All resistance levels correspond to Fig. \ref{rstates}.} 
\vspace{-1em}
\label{vpds2}
\end{figure}

We were able to tune the pixel operating DR, by pre-programming the OxRAM device to specific initial resistance states pre-exposure, for both case(i) and case(ii) switching events. To experimentally validate the achievable resistance states using 10 nm thick HfOx devices, a 4 Kb matrix of 1T-1R OxRAM devices was fabricated and characterized. The obtained $R_{SET}$ and $R_{RESET}$ distributions along with SET and RESET programming conditions are shown in Fig.~\ref{rstates}(a) \& (b) respectively. 
For case (i), Fig.~\ref{rstates}(a) shows six different experimentally programmed R$_{SET}$. For each R$_{SET}$ the devices were first RESET using a fixed RESET programming condition.  R$_{SET}$ programming was achieved by modulating the gate control (WL) voltage in 1T-1R configuration. Physically, $R_{SET}$ control is obtained by controlling the compliance current of the OxRAM device. Compliance current determines the maximum radial dimensions of the conductive filament ~\cite{wong}.

Fig.~\ref{rstates}(b) corresponds to the initialization of case(ii). A SET pulse was applied before application of each RESET condition. In Fig.~\ref{rstates}(b), R$_{RESET}$ values
for write voltage (BL) above 2.5 V saturates to 5 M$\Omega$, due to limitation of the measurement setup used for reading. In reality the R$_{RESET}$
values are much higher, in few $\sim$ G$\Omega$ ~\cite{si_hf1.25n,gaa_hf200p}. 

Fig.~\ref{vpds2}(a) shows the simulated VPD evolution corresponding to four different R$_{SET}$ pre-programming states of OxRAM in pixel, at a fixed exposure of 1 nA. Notice
that for R$_{SET1}$ to R$_{SET3}$, the VPD evolution is almost linear and
non-abrupt. It was observed that even for identical exposure (1 nA), the difference in VPD level at the end of the exposure (10 $\mu$s) can be tuned in a range of 30-60 mV by changing the R$_{SET}$, thus tuning the operating DR.
However for R$_{SET4}$ and lower (i.e. resistance values \textless 10 k$\Omega$), VPD falls instantaneously as no switching occurs in these cases. Switching fails in R$_{SET4}$ and lower than that, as the initial $R_{SET}$ is very low. Low $R_{SET}$ implies larger filament dimensions and need of a stronger corresponding RESET pulse to switch the device. In the inset of Fig.~\ref{vpds2}(a),
we show that by changing the initial VPD level to 1.8 V even low $R_{SET}$ values (\textless 
10 k$\Omega$) can be exploited for the proposed application. Fig.~\ref{vpds2}(b) shows the simulated VPD evolution corresponding to R$_{RESET1}$ and R$_{RESET2}$. Corresponding to each pre-programmed $R_{RESET}$, the VPD evolutions have a difference of $\sim$ 12 mV, for identical exposure (1 nA).

\section{Discussion and Analysis}

\subsection{DR Estimation of Pixel without OxRAM}

For all simulations presented in the paper, a PD full well capacity (FWC) of $\sim$ 62500 \(e-\), an area of $\sim$ 8 $\mu$$m^{2}$, capacitance of $\sim$ 10 fF, with an estimated reset noise~\cite{noise_ana} of $\sim$ 28 \(e-\), were used. 
Pixel operating DR was calculated based on the ratio of maximum to minimum readable exposure as expressed in (2). For the 3T pixel (without OxRAM block, Fig. 3(b)), an operating DR of $\sim$ 20 dB was obtained.  

\begin{equation}
Operating\hspace{0.1cm}DR = 20 log(\frac{I_{exp-max}}{I_{exp-min}})
\end{equation}

In this case, VPD solely corresponds to Iexp as shown in expression (3). 

\begin{equation}
VPD = Vrst- \frac{I_{exp}\hspace{0.1cm}T_{int}}{C_{PD}}
\end{equation}
where $Vrst$ is the PD reset voltage, $C_{PD}$ is the PD capacitance and $T_{int}$ is the charge integration or exposure period.

\subsection{Hybrid Pixel DR Estimation}

In bybrid pixel architecture, VPD evolution also depends on the conduction through the OxRAM device (OxRAM's CF modulation), as expressed in (4).

\begin{equation}
VPD = Vrst - \frac{(I_{exp}+I_{OxRAM})\hspace{0.1cm}T_{int}}{C_{PD}+C_{POX}}
\end{equation}
Where, $C_{POX}$ is the parasitic capacitance of OxRAM due to its MIM type structure, and $I_{OxRAM}$ is current through OxRAM, estimated using (5), (6), and (7) ~\cite{wong2D}.
\begin{equation}
I_{OxRAM} = I_{CF} + I_{oxide}
\end{equation}
\begin{equation}
I_{CF} \hspace{.3cm}  \alpha \hspace{.3cm} exp(-a\hspace{0.1cm}L)  sinh(b\hspace{0.1cm}VPD) 
\end{equation}
\begin{equation}
I_{oxide}  \hspace{.3cm} \alpha \hspace{.3cm}  exp(-c\hspace{0.1cm}x)  sinh(d\hspace{0.1cm} Vgap)
\end{equation}
Where, $I_{CF}$ is the current through the conductive filament of OxRAM, and $I_{oxide}$ is the current through rest of the oxide. $L$ is the thickness of oxide switching layer, $x$ is the length of ruptured CF, and Vgap is the voltage drop in the CF ruptured region. $a$, $b$, $c$, $d$ are OxRAM fixed parameters and constants.   

For case (i), i.e. OxRAM switching from a pre-programed SET-state to RESET-state during the exposure, best case relative DR enhancement of $\sim$ 40 dB was obtained. Corresponding readable exposure I$_{exp}$ was in the range 2.5 pA to 2.5 nA. For case (ii), i.e. OxRAM switching from a pre-programmed partial-RESET state to hard-RESET state during the exposure, best case relative DR enhancement of $\sim$ 50 dB was obtained. Corresponding readable I$_{exp}$ was in the range 0.5 pA to 1.75 nA. In case (iii), i.e. OxRAM switching from pre-programmed RESET-state to SET-state, VPD signal deteriorates very strongly, due to formation of CF. In fact any resistance modulation where the CF is formed or strengthened during the exposure will lead to abrupt fall in VPD.  
The effect of OxRAM's intrinsic capacitance ($C_{POX}$) on the proposed pixel can be deduced from equation (4). In our previous work ~\cite{ash}, we have presented a detailed analysis of OXRAM capacitance calculation and its exploitation for the pixel performance improvement in bright light stimuli. However for the resistive use case described in this paper (with dim-light stimuli), $C_{POX}$ actually degrades the improvement factor of the pixel DR. Higher value of $C_{POX}$ leads to a low charge to voltage conversion gain or low pixel sensitivity. Thus, lower $C_{POX}$ is desirable for the proposed case.

Tab. 1 sumarizes the performance of pixel without OxRAM and pixel with OxRAM for aforementioned switching cases. 
From Tab. 1, in both cases with OxRAM the value of I$_{exp-min}$ is much lower than that for the case without OxRAM. Thus our proposed solution responds much better to low- or dim-lighting situations compared to the pixel without OxRAM. The bright light (I$_{exp-max}$) performance of our solution is marginally decreased compared to pixel without OxRAM, however the relative DR is improved significantly for cases with OxRAM. DR can be further improved by optimizing the pixel operating range and if OxRAM device is engineered to have a higher values of R$_{RESET}$. 

One limiting aspect, on-going/future step of the present study, is to co-integrate/fabricate the pixel and RRAM device in the same die and experimentally validate all simulations discussed. The present paper is a first step in that direction. Also, for the proposed resistive technique, the hybrid pixel shows no DR enhancement for bright/normal light exposure stimuli. However, we have shown a possible solution for bright-light case earlier ~\cite{ash}; by exploiting purely capacitive behavior of OxRAM. 
The technique described in this paper mainly helps with DR enhancement in dim-light stimuli, thus one may assume that for bright- and normal- light stimuli the OxRAM device doesn’t need to actively switch or participate in the pixel circuit, and can be virtually disconnected by switching off its select transistor. This drastically brings down the number of cases where the OxRAM device actually needs to switch, and helps overcome the impact of limited device endurance. Impact of endurance will further depend on the exact nature of the stimuli, the spatial dependence of lighting conditions, and the actual size of the pixel array. In a large pixel array, even for dim light stimuli, it is not necessary that every single pixel has the same low intensity of light falling on it. Thus all OxRAM devices in the pixel array need not switch in every dim-light event. High endurance of the order of $>$10$^{5}$ cycles has been reported in ~\cite{gaa_hf200p} for state-of-the art HfOx devices. Choosing optimized programming conditions may also enhance the endurance of OxRAM devices. Weak SET state programming, like the one useful for the proposed application, further helps to improve the device endurance.
Another limitation of the proposed approach may arise while precise tuning of the operating DR using pre-programming of the OxRAM device. Several OxRAM stacks have been shown to have wide resistance distributions while being programmed in the RESET state, however SET state programming can be achieved with less variability through compliance current and selector devices. Apart from the DR enhancement, the in-pixel OxRAM devices lead to a significant area benefit as they can be fully integrated in back-end or vias between two metal levels.     

\begin{table}
\centering
\caption{Performance comparison of pixel with and without OxRAM}
\begin{tabular}[c]{ |p{1.2cm}|p{1.45cm}|p{1.9cm}%|p{1.3cm}
|p{1.8cm}|}
\hline
Pixel, Switching Type& Minimum Readable\newline Exposure\newline I$_{exp-min}$ \newline (pA) & Maximum Readable\newline Exposure\newline I$_{exp-max}$ \newline (pA) 
& Relative operating DR Improvement \newline(dB)\\
\hline
Without OxRAM & 315 & 3.1 x 10$^{3}$ %& - 
& - \newline ( 20 )\\
\hline
With OxRAM, Case (i)& 2.5 & 2.5 x 10$^{3}$ %& 3.08 
& $\sim$ 40 \newline ( 60 )\\
\hline
With OxRAM, Case (ii)& 0.5 & 1.75 x 10$^{3}$ %& 102.8 
& $\sim$ 50 \newline ( 70 )\\
\hline
\end{tabular}
\label{tab1}
\end{table}

\section{Conclusion}
We show a first of its kind unique method to improve operating dynamic range of CMOS Image Sensor pixels by exploiting HfOx based OXRAM devices. We detailed the working of our hybrid CMOS-OxRAM pixel circuit, and analyzed different preconditioning strategies for the same. We showed that by OxRAM resistance modulation the DR can be tuned. Best case relative DR enhancement of $\sim$ 40 dB, and $\sim$ 50 dB were obtained for SET-to-RESET state, and soft-RESET to hard-RESET state transitions. The use of emerging nonvolatile memory device (OxRAM) in the proposed circuit can be further exploited for in-pixel information storage along with proposed DR improvement technique. The proposed work opens door for an entirely new and exciting non-storage application domain of emerging RRAM devices.

\section*{Acknowledgment} 
The PI Dr. Manan Suri, would like to acknowledge the partial support of DST-Government of India, sponsored research project grant RP03051. We also thank E. Nowak and L. Perniola of CEA-LETI, France, for the fabrication and characterization of the 4Kb 1T-1R OxRAM matrix for this study.

\end{document}